\documentclass[11pt,a4paper]{article}
\usepackage{jcappub,natbib}

\title{Axion mass limit from observations of the neutron star in Cassiopeia A}

\author{Lev B. Leinson}
\affiliation{Pushkov Institute of Terrestrial Magnetism, Ionosphere and Radiowave Propagation of the Russian Academy of Science (IZMIRAN),\\142190 Troitsk, Moscow, Russia}
\emailAdd{leinson@yandex.ru}

\abstract{Direct Chandra observations of a surface temperature of isolated neutron star in Cassiopeia A (Cas A NS) and its cooling scenario which has been recently simultaneously suggested by several scientific teams put stringent constraints on poorly known  properties of the superfluid neutron star core. It was found also that the thermal energy losses from Cas A NS are approximately twice more intensive than it can be explained by the neutrino emission.  We use these unique data and well-defined cooling scenario to estimate the strength of KSVZ axion interactions with neutrons. We speculate that enlarged energy losses occur owing to emission of axions from superfluid core of the neutron star. If the axion and neutrino losses are comparable we find  $c_{n}^{2}m_{a}^{2}\sim 5.7\times 10^{-6}\,\text{eV}^2$, where $m_{a}$ is the axion mass, and  $c_{n}$ is the effective Peccei-Quinn charge of the neutron. (Given the QCD uncertainties of the hadronic axion models, the dimensionless constant $c_{n}$ could range from $-0.05$ to $ 0.14$.)}


\begin{document}
\maketitle
\flushbottom

\section{Introduction}
\label{sec:intro}

Axions are hypothetical pseudoscalar particles that have been suggested as a
solution of the CP-violation problem in the strong interactions \cite%
{P77,W78,Wl78}. Though axions arise as Nambu-Goldstone bosons and thus must
be fundamentally massless their interaction with gluons induces their mixing
with neutral pions. Axions thereby acquire a small mass which is
approximately given by \cite{BT78,KSS78,PBY87,GKR86}: 
\begin{equation}
m_{a}=0.60\,\text{eV}\,\frac{10^{7}\text{GeV}}{f_{a}},  \label{ma}
\end{equation}%
where the unknoun constant $f_{a}$ with the dimension of an energy is the
axion decay constant. For a general review on axion physics see, e.g., \cite%
{KSVZ,KSVZ1,DFSZ,DFSZ1,Kim,Cheng}. The axion phenomenology, in particular in relation
with the astrophysical processes, is largely discussed in \cite%
{VZKC,R,R90,T90,R99,R08}.

Axions are a plausible candidate for the cold dark matter of the universe,
and a reasonable estimate of the axion mass (or, equivalently, the axion
decay constant) represents much interest. Over the years, various laboratory
experiments as well as astrophysical arguments have been used to constrain
the allowed range for $f_{a}$ or, equivalently, for the axion mass $m_{a}$.
Currently \cite{AS83,DF83}, cosmological arguments give $m_{a}>10^{-5}$ eV.
The most stringent upper limits on the axion mass derive from astrophysics. 

Axions produced in hot astrophysical plasma can transport energy out of
stars. Strength of the axion coupling with normal matter and radiation is
bounded by the condition that stellar-evolution lifetimes or energy-loss
rates not conflict with observation. Such arguments are normally applied to
the physics of supernova explosions, where the dominant energy loss process
is the emission of neutrino pairs and axions in the nucleon bremsstrahlung 
\cite{Br88,Bu88,R93,H00}. The limit from Supernova 1987A gives $m_{a}<0.01$
eV \cite{RS91,J96}. In works \cite{I84,U97} the thermal evolution of a
cooling neutron star was studied by including the axion emission in addition
to neutrino energy losses. The authors suggest the
upper limits on the axion mass of order $m_{a}<0.06-0.3$ eV by comparing the
theoretical curves with the ROSAT observational data for three pulsars: PSR
1055-52, Geminga and PSR 0656+14. Accuracy of such estimates substantially
depends on the assumptions of the matter equation of state and of the
effects of nucleon superfluidity which should be properly taken into
account. In the most cases the cooling scenario involves many parameters
which are poorly known.

The possibility of a more correct estimate has appeared following a 
publication of analysed Chandra observations of the  neutron star in 
Cassiopeia A (Cas A NS) during 10 years \cite{H09,H10}. The authors found a 
steady decline of the surface temperature, $T_{s}$, by about 4\% which they 
interpret as a direct observation of Cas A NS cooling, the phenomenon which 
has never been observed before for any isolated NS. 
The decline is naturally explained if
neutrons have recently become superfluid (in $^{3}$P$_{2}$ triplet-state) in
the NS core, producing a splash of neutrino from pair breaking and formation 
(PBF) processes\footnote{In Ref. \cite{St} the authors use the term Cooper 
pair formation (CPF).} that currently accelerates the cooling \cite{St,P}. 
The observed rapidity of the Cas A NS cooling implies that protons were 
already in a superconducting $^{1}$S$_{0}$ singlet-state with a larger 
critical temperature.
This scenario puts stringent constraints on poorly known properties of NS
cores.  In particular, the density dependence of the temperature for the
onset of neutron superfluidity should have a wide peak with maximum $%
T_{c}(\rho )\approx (7-9)\times 10^{8}$~K.

\section{Neutrino and axion energy losses from superfluid NS core}
\label{sec:emis}

The neutrino pair emission caused by recombination of thermally broken Cooper
pairs \cite{FRS76,YKL} occurs through neutral weak currents generated by spin
fluctuations of the nucleons \cite{LP06,L10}. 
Since the proton condensation occurs with a zeroth total spin
of a Cooper pair the spin fluctuations of the proton condensate are strongly
suppressed in the non-relativistic system \cite{FRS76}. As a result, the
dominating energy losses occur owing to the PBF neutrino radiation from
triplet pairing of neutrons, while the proton superfluidity quenches the
other neutrino reactions which efficiently operate in normal (nonsuperfluid)
nucleonic systems ($\bar{\nu}\nu $ bremsstrahlung, murca processes etc.)

Since the neutrino emission occurs mainly owing to neutron spin
fluctuations, the part of the interaction Hamiltonian relevant for PBF
processes is (we use natural units, $\hbar =c=k_{B}=1$):%
\begin{equation}
\mathcal{H}_{\nu n}=-\frac{G_{F}C_{\mathsf{A}}}{2\sqrt{2}}\delta
_{\mu i}\left( \Psi ^{+}\hat{\sigma}_{i}\Psi \right)l^{\mu } ,  \label{Hnu}
\end{equation}%
where $l^{\mu }=\bar{\nu}\gamma ^{\mu }\left( 1-\gamma _{5}\right) \nu $ is
the neutrino current, $G_{F}=1.166\times 10^{-5}$ GeV$^{-2}$ is the Fermi
coupling constant, $\Psi $ is the nucleon field, $C_{\mathsf{A}}\simeq
1.26$ is the neutral-current axial-vector coupling constant of neutrons, and 
$\hat{\sigma}_{i}$ are the Pauli spin matrices.

The dominant axion emission from a hot neutron star core is also caused by 
spin fluctuations of non-relativistic neutrons. The corresponding Hamiltonian
density can be written in the form of derivative coupling:%
\begin{equation}
\mathcal{H}_{an}=\frac{c_{n}}{2f_{a}}\delta _{\mu i}\left( \Psi ^{+}\hat{%
\sigma}_{i}\Psi \right) \partial ^{\mu }a,  \label{Ha}
\end{equation}%
where $c_{n}$ is the effective Peccei-Quinn charge of the neutron. This 
dimensionless, 
model-dependent coupling constant could range from $-0.05$ to $ 0.14$
\cite{M88,M89}.

The emission of neutrino pairs is kinematically possible owing to the
existence of a superfluid energy gap, which admits the quasiparticle
transitions with time-like momentum transfer $K=\left( \omega ,\mathbf{k}%
\right) $, as required by the final neutrino pair: $K=K_{1}+K_{2}$. \ The
energy-loss rate by $\bar{\nu}\nu $ emission caused by the neutron PBF
processes is given by the phase-space integral 
\begin{equation}
Q_{\bar{\nu}\nu }\simeq \mathcal{N}_{\nu }\frac{G_{F}^{2}C_{\mathsf{A}}^{2}}{%
8}\int \;\frac{\omega }{1-\exp \frac{\omega }{T}}2\operatorname{Im}\Pi _{%
\mathsf{A}}^{\mu \nu }\left( \omega \right) \mathrm{Tr}\left( l_{\mu }l_{\nu
}^{\ast }\right) \frac{d^{3}k_{1}}{2\omega _{1}(2\pi )^{3}}\frac{d^{3}k_{2}}{%
2\omega _{2}(2\pi )^{3}},  \label{Q}
\end{equation}%
where $\mathcal{N}_{\nu }=3$ is the number of neutrino flavors, and 
$\Pi _{\mathsf{A}}^{\mu \nu }$ is the retarded axial polarization tensor which 
describes spin fluctuations in the neutron superfluid at temperature $T$. 
The Fermi velocity is small in the nonrelativistic system, $V_{F}\ll 1$, 
and we can study the neutrino energy losses in the lowest order over this small
parameter. Since the transferred space momentum comes in the polarization
functions in a combination $\mathbf{kV}_{F}\ll \omega ,\Delta $, one can
evaluate $\Pi _{\mathsf{A}}^{\mu \nu }$ in the limit $\mathbf{k}=0$.

After integration over the phase space of escaping neutrinos and
antineutrinos the total energy which is emitted into neutrino pairs per unit
volume and time is given by the following formula (See details, e.g., in
Ref. \cite{L01}): 
\begin{equation}
Q_{\bar{\nu}\nu }=\frac{G_{F}^{2}C_{\mathsf{A}}^{2}}{64\pi ^{5}}%
\int_{0}^{\infty }d\omega \int\limits_{k<\omega }d^{3}q\frac{\omega }{1-\exp 
\frac{\omega }{T}}\operatorname{Im}\Pi _{\mathsf{A}}^{\mu \nu }\left( \omega \right)
\left( K_{\mu }K_{\nu }-K^{2}g_{\mu \nu }\right) ~,  \label{QQQ}
\end{equation}%
where we use a shortened notation $\Pi _{\mathsf{A}}^{\mu \nu }\left( \omega
\right) \equiv \Pi _{\mathsf{A}}^{\mu \nu }\left( \omega ,\mathbf{k}%
=0\right) $.

If now $K=\left(k,\mathbf{k}\right) $ denotes the axion four-momentum (we
ignore a small axion mass), the energy radiated per unit volume and  time in
axions is given by the following phase-space integral 
\begin{equation}
Q_{a}=\frac{1}{4}\frac{c_{n}^{2}}{f_{a}^{2}}\int \;\frac{k}{1-\exp \frac{k}{T%
}}2\operatorname{Im}\Pi _{\mathsf{A}}^{\mu \nu }\left( k\right) K_{\mu }K_{\nu }%
\frac{d^{3}k}{2k(2\pi )^{3}}.  \label{rate1}
\end{equation}%
In the above, it was assumed that both axions and neutrinos can escape
freely from the medium so that final-state Pauli blocking factors can be
ignored.

The medium properties are embodied in a common function $\operatorname{Im}\Pi _{%
\mathsf{A}}^{\mu \nu }$ which is exactly the same for axion or neutrino
interactions because in Eqs. (\ref{QQQ}) and (\ref{rate1}) the global
coupling constants are explicitly pulled out. For the $^{3}$P$_{2}(m_{j}=0)$
pairing of neutrons this function is calculated in Ref. \cite{L10} with
taking into account of the ordinary and anomalous axial-vector vertices.
According to Eq. (93) of this work:%
\begin{align}
\operatorname{Im}\Pi _{\mathsf{A}}^{\mu \nu }\left( \omega \right) & =-\delta ^{\mu
i}\delta ^{\nu j}\frac{p_{F}M^{\ast }}{\pi ^{2}}\int d\mathbf{n}\left(
\delta _{ij}-\frac{\bar{b}_{i}\bar{b}_{j}}{\bar{b}^{2}}-\frac{3}{4}\left(
\delta _{ij}-\delta _{i3}\delta _{j3}\right) \right)  \notag \\
& \times \Delta _{\mathbf{n}}^{2}\frac{\Theta \left( \omega ^{2}-4\Delta _{%
\mathbf{n}}^{2}\right) }{\omega \sqrt{\omega ^{2}-4\Delta _{\mathbf{n}}^{2}}}%
\tanh \frac{\omega }{4T},  \label{ImPi}
\end{align}%
where $p_{F}$ is the Fermi momentum of neutrons, $M^{\ast }\equiv
p_{F}/V_{F}$ is the neutron effective mass, and $\Theta \left( x\right) $ is the 
Heaviside step-function.  
For the $^{3}$P$_{2}(m_{j}=0)$ pairing the
normalised vector $\mathbf{\bar{b}}\left( \mathbf{n}\right) $ is defined as 
\begin{equation}
\mathbf{\bar{b}}\left( \mathbf{n}\right) \equiv \sqrt{1/2}\left(
-n_{1},-n_{2},2n_{3}\right) .  \label{bb}
\end{equation}%
Its angular dependence is represented by the unit vector $\mathbf{n=p}/p$
which defines the polar angles $\left( \theta ,\varphi \right) $ on the
Fermi surface:%
\begin{equation}
\mathbf{n=}\left( \sin \theta \cos \varphi ,\sin \theta \sin \varphi ,\cos
\theta \right) .  \label{nn}
\end{equation}
The superfluid energy gap, generally defined by the relation  
\begin{equation}
\Delta _{\mathbf{n}}^{2}=\mathbf{\bar{b}}^{2}\left( \mathbf{n}\right) \
\Delta ^{2}\left( \tau \right) ,  \label{Dns}
\end{equation}%
is anisotropic. It depends on the polar angle $\theta $ and on the relative
temperature $\tau \equiv T/T_{c}$. For the one component state $m_{j}=0$ one
has 
\begin{equation}
\Delta _{\mathbf{n}}=\frac{1}{\sqrt{2}}\sqrt{1+3\cos ^{2}\theta }\ \Delta
\left( \tau \right) .  \label{Dn}
\end{equation}

Insertion of Eq. (\ref{ImPi}) into Eqs. (\ref{QQQ}) and (\ref{rate1}) yields
the neutrino emissivity as given by Eq. (96) of Ref. \cite{L10}:

\begin{equation}
Q_{\nu }(m_{j}=0)\simeq \frac{2}{5\pi ^{5}}G_{F}^{2}C_{\mathsf{A}%
}^{2}p_{F}M^{\ast }T^{7}F_{4}\left( \frac{T}{T_{c}}\right) ~,
\label{Qnu}
\end{equation}%
and the axion emissivity 
\begin{equation}
Q_{a}(m_{j}=0)=\frac{c_{n}^{2}}{f_{a}^{2}}\frac{2}{3\pi ^{3}}p_{F}M^{\ast
}T^{5}F_{2}\left( \frac{T}{T_{c}}\right) ,  \label{Qa}
\end{equation}%
where%
\begin{equation}
F_{l}\left( \tau \right) =\int \frac{d\mathbf{n}}{4\pi }\frac{%
\Delta _{\mathbf{n}}^{2}}{T^{2}}\int_{0}^{\infty }dx\frac{z^{l}}{\left( \exp
z+1\right) ^{2}}  \label{J}
\end{equation}%
with $z=\sqrt{x^{2}+\Delta _{\mathbf{n}}^{2}/T^{2}}$. Details of the
numerical evaluation of this integral can be found in \cite{YKL,L10}.

\section{Mixed cooling by emission of axions and neutrino pairs}
\label{sec:mix}

Before proceeding to estimates of the axion radiation, let us note a few important 
details of theoretical simulation of the CAS A NS neutrino cooling.
The authors of Ref. \cite{St} have reported that our Eq. (\ref{Qnu}) gives too slow 
cooling. To achieve a better quantitative agreement of their simulation to the 
observed data the neutrino energy losses were artifically enlarged in approximately 
two times. This indicates that the thermal energy losses of Cas A NS are 
approximately twice more intensive than neutrino losses given in Eq. (\ref{Qnu}).
Since currently there is no definitive explanation for this increase, we can speculate 
that the additional energy losses from the superfluid core of the Cas A NS are 
caused by axion emission, as described in Eq. (\ref{Qa}).

To get an idea of a compatibility of the axion emission with the CAS A NS observation 
data let 
us consider a simple model of cooling of the superfluid neutron core enclosed in a 
thin envelope as typical for the NS. We assume that the bulk matter consists mostly 
of $^{3}P_{2}$ superfluid neutrons with $m_{j}=0$. 
In the temperature range which we are interested in, the thermal 
luminosity of the surface is negligible in comparison to neutrino and 
axion luminosities of the PBF processes in the NS core. In this case 
the equation of global thermal balance \cite{gs80} reduces to
\begin{equation}
   C(\widetilde{T}) \, { {\rm d} \widetilde{T} \over {\rm d} t}=
   -L (\widetilde{T}).
\label{aa}
\end{equation}
Here $L (\widetilde{T})$ is the total 
PBF luminosity of the star (redshifted to a distant observer), while 
$C(\widetilde{T})$ is the stellar heat capacity. These quantities 
are given by (see details in Refs. \cite{Y}):
\begin{eqnarray}
   L (\widetilde{T})& = & \int {\rm d}V \, Q(T,\rho)
\exp(2 \Phi(r)),
\label{eq:Lnu} \\
 C(\widetilde{T})& = & \int {\rm d}V \, C_V(T,\rho),
\label{C}
\end{eqnarray}
where $Q(T,\rho)$ is the total  (neutrino + axion) emissivity,
$C_V (T,\rho)$ is the specific heat capacity,
${\rm d}V$ is the element of proper volume determined by
the appropriate metric function, and $\Phi(r)$ is the
metric function that determines gravitational redshift.
A thermally 
relaxed star has an isothermal interior which extends from
the center to the heat blanketing envelope.
Taking into account the effects of General
Relativity (e.g., \cite{thorne77}), isothermality means
spatially constant redshifted internal temperature
\begin{equation}
\widetilde{T}(t)=T(r,t) \exp(\Phi(r)), \label{bb}
\end{equation}
while the local internal temperature $T(r,t)$, 
depends on the radial coordinate $r$.

Given the strong dependence of the PBF processes on the temperature and the 
density, the overall effect of simultaneous emission of neutrino pairs and axions can 
only be assessed by complete calculations of the neutron star cooling which are
beyond the scope of this paper. A rough estimate can be made in a simplified model, 
where the superfluid transition temperature $T_c$ is constant over the core. 

In the temperature range of our interest, the specific heat is governed by 
the neutron component (the contribution of electrons and strongly superfluid 
protons is negligibly small) and can be described as
\begin{equation}
 C\simeq\frac{1}{3}TR_{B}(T/T_c)
\int dVp_{F}M^{\ast }, \label{cc}
\end{equation}
 where $R_B(T/T_c)$ is the superfluid reduction factor, as given in Eq. (18) of 
Ref. \cite{YLS}.  

Making use of Eqs. (\ref{Qnu}) and (\ref{Qa}) we obtain the PBF luminosity in 
the form
\begin{equation}
L =\left[\frac{2}{5\pi ^{5}}G_{F}^{2}C_{%
\mathsf{A}}^{2}T^{7}F_{4}(T/T_c) +
\frac{c_{n}^{2}}{f_{a}^{2}}\frac{2}{3\pi ^{3}}%
T^{5}F_{2}(T/T_c)\right] \int dVp_{F}M^{\ast }e^{2\Phi ( r) }. 
\label{dd}
\end{equation}
Insertion of  Eqs. (\ref{bb}), (\ref{cc}) and (\ref{dd}) into Eq. (\ref{aa}) allows to obtain the 
following equation for the non-redshifted temperature $T_{b}(t)\equiv T(r_b,t)$ at the edge 
of the core or, equivalently, at the bottom of the envelope at $r=r_b$:
\begin{equation}
\frac{dT_{b}}{dt}=\frac{3\alpha }{R_{B}\left( T_{b}/T_{c}\right) }\left[ \frac{2%
}{5\pi ^{5}}G_{F}^{2}C_{\mathsf{A}}^{2}T_{b}^{6}F_{4}\left(
T_{b}/T_{c}\right) +\frac{c_{n}^{2}}{f_{a}^{2}}\frac{2}{3\pi ^{3}}%
T_{b}^{4}F_{2}\left( T_{b}/T_{c}\right) \right], \label{Tbeq} 
\end{equation}
where the constant $\alpha\equiv\alpha (r_b)$ is defined as
\begin{equation}
\alpha \equiv \frac{\int dVp_{F}M^{\ast }e^{2\Phi \left( r\right) }}{\exp
\Phi \left( r_{b}\right) \int dVp_{F}M^{\ast }}, 
\label{alpha}
\end{equation}
and can be found from the CAS A NS observation data. 

We convert the internal $T_b$ to the
observed effective surface temperature $T_{s}$ using (see, e.g., \cite{GPE,p06}) 
\begin{equation}
T_{s}/10^{6}\mathrm{K}\simeq (T_{b}/10^{8}\mathrm{K})^{0.55 }.
\label{Eq:TeTb}
\end{equation}%
This allows to compare the computed results with the observed (non-redshifted) CAS A NS 
surface temperatures which are cataloged in Table 1 of Ref. \cite{St}.

\section{Results and discussion}
\label{sec:result}

\begin{figure}
  \begin{center}
    \includegraphics[width =1\textwidth]{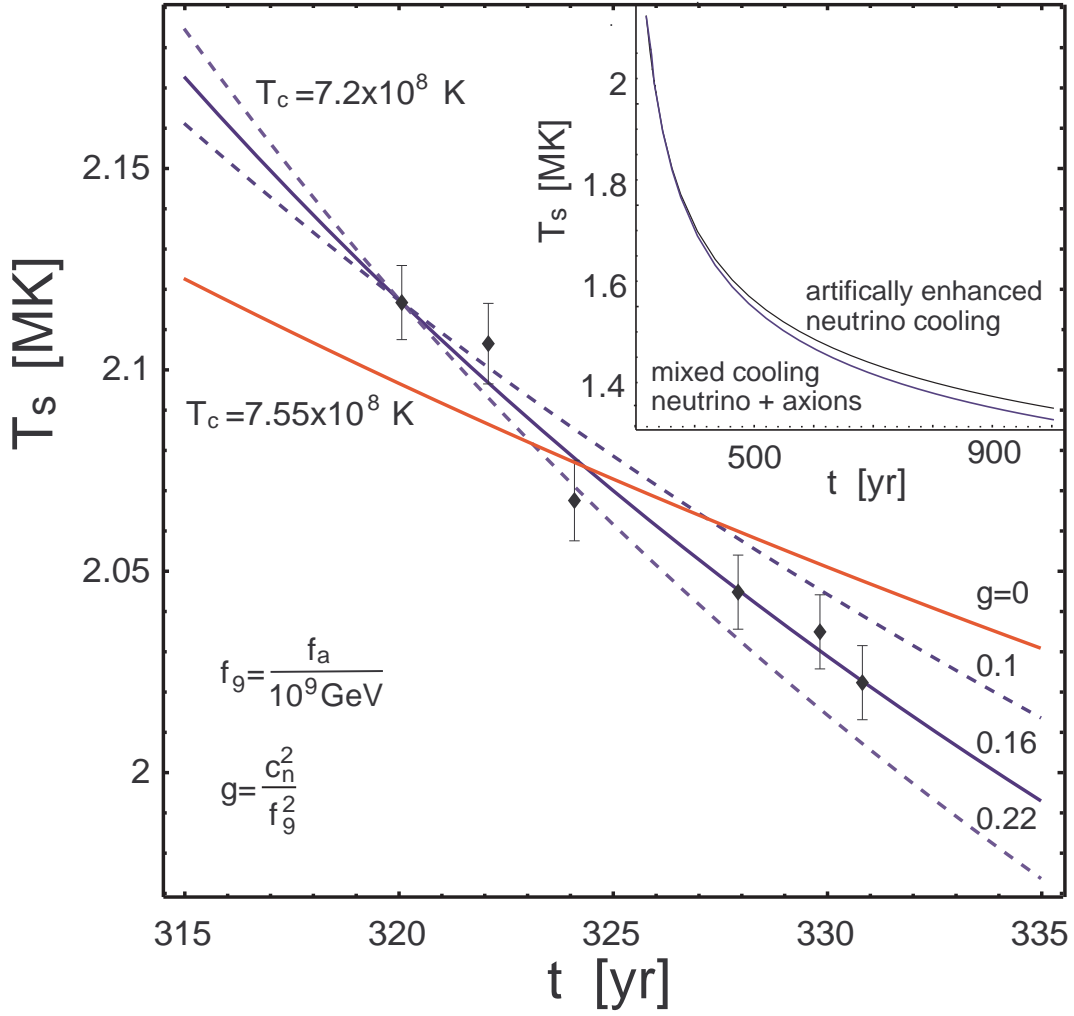}
  \end{center}
  \caption{(Color on line)
Cooling curves for a simulated CAS A NS consisting of a superfluid neutron core 
and a low-mass blanketing envelope. $T_c$ is taken constant over the core. 
Four curves correspond to the mixed (neutrino + axion) cooling at four values 
$g=0$ ($T_c=7.55\times 10^8$~K), $g=0.1,\ 0.16$ and 
$0.22$ ($T_c=7.2\times 10^8$~K). The points with error bars demonstrate 
the observed surface temperatures cataloged in Table 1 of Ref. \cite{St}.
The inset shows the cooling curves but over larger range of ages. The lower 
curve corresponds to the mixed cooling at  $g=0.16$ while the upper curve 
demonstrates cooling due to only neutrino emission artifically enhanced 2.1 times 
as suggested in Ref. \cite{St}.}
  \label{fig:fig1}
\end{figure}

For numerical estimate of the axion coupling strength to neutrons we designate
\begin{equation}
g=\frac{c_{n}^{2}}{f_{9}^{2}}, \label{g}
\end{equation}
with $f_{9}=f_{a}/\left( 10^{9}\text{GeV}\right)$, and consider $g$ as a free 
parameter.  Fig. \ref{fig:fig1} demonstrates the effect of mixed cooling of superfluid 
neutron star with a constant $T_c$ over the core. 
Two solid lines are the cooling curves for the simulated NS calculated at 
$g=0.0$ and $g=0.16$. 
The case $g=0$ describes the cooling caused by only the PBF neutrino emission 
given in Eq. (\ref{Qnu}), with constant $T_c=7.55\times 10^8$ K. This curve 
demonstrates too slow cooling and cannot explain the data. The case 
$g=0.16$ agrees with the observations.  This corresponds to the mixed 
neutrino + axion radiation, as described by Eqs. (\ref{Qnu}) and (\ref{Qa}), 
with  $T_c=7.2\times 10^8$ K. The two dashed curves calculated at $g=0.1$ and 
$g=0.22$ demonstrate that even a relatively small deviation off the value 
$g=0.16$ results in substantial modification of the temperature profile and does 
not allow to reproduce the observed cooling rate of the Cas A NS. 

Thus we obtain $g\simeq 0.16$ or, equivalently,
\begin{equation}
\frac{c_{n}^{2}}{f_{a}^{2}}\simeq 1.6\times 10^{-19} \text{GeV}^{-2}.
\label{fa}
\end{equation}.

The same estimate immedeatly follows from a simple comparison of Eqs. 
(\ref{Qnu}) and (\ref{Qa}) if one assumes that the axionic emissivity is 
approximately equal to the neutrino emissivity. This gives
\begin{equation}
\frac{c_{n}^{2}}{f_{a}^{2}}\sim \frac{3}{5\pi ^{2}}G_{F}^{2}C_{\mathsf{A}%
}^{2}T^{2}\frac{F_{4}\left( \frac{T}{T_{c}}\right) }{F_{2}\left( \frac{T%
}{T_{c}}\right) }.  \label{R1}
\end{equation}%
Inserting the typical values, $T_{c}\simeq 7.2\times 10^{8}\mathsf{K}$ and 
$T\simeq 3.8\times 10^{8}\mathsf{K}$, we find $\tau \equiv T/T_{c}\simeq 0.53$ and
$F_{4}\left( \tau\right) /F_{2}\left( \tau \right) \simeq 10.4$. 
Insertion of the above parameters into Eq (\ref{R1}) results in the estimate 
given in Eq. (\ref{fa}).

 One can use Eq. (\ref{ma}) to convert the decay constant $f_{a}$ to the axion mass 
$m_{a}$. This yields
\begin{equation}
 c_{n}^{2}m_{a}^{2}\sim 5.7\times 10^{-6}\,\text{eV}^2.  \label{mc}
\end{equation}%

Unfortunately, the coupling constant $c_{n}$ depends on the axion model. 
Given the QCD uncertainties of the hadronic axion models \cite{K85,S85,G82}, the 
dimensionless constant $c_{n}$ could range from $-0.05$ to $ 0.14$.
While the canonical value $c_{n}=0.044$ is often used as generic examples, in 
general $c_{n}$ is not known so that for fixed $ c_{n}^{2}m_{a}^{2}$ a broad 
range of $m_a$ values is possible. 

One should keep in mind that a strong cancelation of $c_{n}$ below 
$c_{n}=0.044$ is also allowed. 
In case of $c_{n}\rightarrow 0$ a powerfull PBF emission of axions is impossible. 
This would mean that our assumption of the mixed cooling is invalid, and the PBF 
neutrino losses are indeed at least two times larger than is predicted in Eq. 
(\ref{Qnu}). Then the axion energy losses produce no noticeable modification of the 
temperature profile of the CAS A NS, and one has to replace the Eq. (\ref{mc}) by 
the inequality
\begin{equation}
 c_{n}^{2}m_{a}^{2}\ll 5.7\times 10^{-6}\,\text{eV}^2.  \label{mcl}
\end{equation}%

Can we discriminate the two cases from observations of the NS surface temperature? 
As demonstrated in the insert in Fig. 1, the difference between the corresponding 
theoretical cooling curves becomes discernable only after about 1000 years of cooling. 
Perhaps future observation of the surface temperature of old neutron stars will help 
to clarify the cooling mechanism.

Finally let us notice that our estimate of interaction of the hadronic axions with neutrons 
has no analogies for a comparison. Previous astrophysical constraints was derived basically 
for axions interacting simultaneously with neutrons and protons. In our case the proton 
contribution is turned off due to large superfluid energy gap.


\begin{thebibliography}{99}
\bibitem{P77} R.D. Peccei and H.R. Quinn, \emph{CP Conservation In The Presence Of Instantons}, 
\emph{Phys. Rev. Lett.} {\bf 38} (1977) 1440.

\bibitem{W78} S. Weinberg, \emph{A New Light Boson?}, \emph{Phys. Rev. Lett.} {\bf 40} (1978) 223.

\bibitem{Wl78} F. Wilczek, \emph{Problem Of Strong P And T Invariance In The Presence Of Instantons}, 
\emph{Phys. Rev. Lett.} {\bf 40} (1978) 279.

\bibitem{BT78} W.A. Bardeen and S.H.H. Tye, \emph{Current Algebra Applied to Properties of the Light Higgs Boson}, \emph{Phys. Lett.} {\bf B 74} (1978) 229.

\bibitem{KSS78} J. Kandaswamy, P. Salomonson and J. Schechter, \emph{Mass of the Axion}, \emph{Phys. Rev.} {\bf D 17} (1978) 3051.

\bibitem{PBY87} W.A. Bardeen, R.D. Peccei and T.Yanagida, \emph{Constraints On Variant Axion Models}, \emph{Nucl. Phys.} {\bf B 279} (1987) 401.

\bibitem{GKR86} H. Georgi, D.B. Kaplan and L.Randall, \emph{Manifesting the Invisible Axion at Low-energies}, \emph{Phys. Lett.} {\bf B 169} (1986) 73.

\bibitem{KSVZ} J.E. Kim, \emph{Weak Interaction Singlet and Strong CP Invariance}, \emph{Phys. Rev. Lett.} {\bf 43} (1979) 103.

\bibitem{KSVZ1} M.A. Shifman, A.I. Vainshtein and V.I. Zakharov, \emph{Can Confinement Ensure Natural CP Invariance Of Strong Interactions?}, \emph{Nucl. Phys.} {\bf B 166} (1980) 493.

\bibitem{DFSZ} A.R. Zhitnitskii, \emph{Possible suppression of axion-hadron interactions}, \emph{Sov. J. Nucl. Phys.} {\bf 31} (1980) 260.

\bibitem{DFSZ1} M. Dine, W. Fischler and M. Srednicki, \emph{A Simple Solution To The Strong CP Problem With A Harmless Axion}, \emph{Phys. Lett.} {\bf B 104} (1981) 199.

\bibitem{Kim} J.E. Kim, \emph{Light Pseudoscalars, Particle Physics And Cosmology}, \emph{Phys. Rept.} {\bf 150} (1987) 1.

\bibitem{Cheng} H.Y. Cheng, \emph{The Strong CP Problem Revisited}, \emph{Phys. Rept.} {\bf 158} (1988) 1.

\bibitem{VZKC} M.I. Visotsskii, Ya.B. Zel'dovich, M.Yu. Khlopov and V.M. Chechetkin, \emph{Some astrophysical limitations on axion mass} \emph{Pis'ma v ZhETF} {\bf 27} (1978) 533; [English translation: \emph{JETP Lett.} {\bf 27} (1978) 502].

\bibitem{R} G.G. Raffelt, \emph{Stars as Laboratories for Fundamental Physics.} \emph{The Astrophysics of Neutrinos, Axions, and Other Weakly Interacting Particles.}, The University of Chicago Press, U.S.A. (1996).

\bibitem{R90} G.G. Raffelt, \emph{Astrophysical methods to constrain axions and other novel particle phenomena}, 
\emph{Phys. Rept.} {\bf 198} (1990) 1.

\bibitem{T90} M.S. Turner, \emph{Windows On The Axion}, \emph{Phys. Rept.} {\bf 197} (1990) 67.

\bibitem{R99} G.G. Raffelt, \emph{Particle physics from stars}, \emph{Ann.Rev.Nucl.Part.Sci.} {\bf 49} (1999) 163.

\bibitem{R08} G.G. Raffelt, \emph{Astrophysical axion bounds}, \emph{Lct. Notes Phys.} {\bf 741} (2008) 51.

\bibitem{AS83} L.F. Abbott and P.Sikivie, \emph{A Cosmological Bound on the Invisible Axion}, \emph{Phys. Lett.} {\bf B 120} (1983) 133.

\bibitem{DF83} M. Dine and W. Fischler, \emph{The Not So Harmless Axion}, \emph{Phys. Lett} {\bf B 120} (1983) 137.

\bibitem{Br88} R.P. Brinkmann and M.S. Turner, \emph{Numerical Rates for Nucleon-Nucleon Axion Bremsstrahlung}, 
\emph{Phys. Rev.} {\bf D 38} (1988) 2338.

\bibitem{Bu88} A. Burrows, M.S. Turner and R.P. Brinkmann, 
\emph{Axions and SN 1987a}, \emph{Phys. Rev.} {\bf D 39} (1989) 1020.

\bibitem{R93} G. Raffelt and D. Seckel, 
\emph{A selfconsistent approach to neutral current processes in supernova cores}, 
\emph{Phys. Rev.} {\bf D 52} (1995) 1780.

\bibitem{H00} C. Hanhart, D.R. Phillips and S. Reddy, 
\emph{Neutrino and axion emissivities of neutron stars from nucleon-nucleon scattering data},
\emph{Phys. Lett.} {\bf B 499} (2001) 9.

\bibitem{RS91} G. Raffelt and D. Seckel, \emph{Multiple scattering suppression of the bremsstrahlung emission of neutrinos and axions in supernovae}, \emph{Phys. Rev. Lett.} {\bf 67} (1991) 2605.

\bibitem{J96} H.-T. Janka, W. Keil, G. Raffelt and D. Seckel, \emph{Nucleon spin fluctuations and the supernova emission of neutrinos and axions}, \emph{Phys. Rev.Lett.} {\bf 76} (1996) 2621.

\bibitem{I84} N. Iwamoto, \emph{Axion Emission from Neutron Stars}, \emph{Phys. Rev. Lett.} {\bf 53} (1984) 1198.

\bibitem{U97} H. Umeda, N. Iwamoto, S. Tsuruta, L. Qin and K. Nomoto, \emph{Axion mass limits from cooling neutron stars},
in Proceedings of the International Conference on Neutron Stars and Pulsars, edited by N. Shibazaki et al., Universal Academy Press (Frontiers science series no. 24), p. 213 (1998) [astro-ph/9806337].

\bibitem{H09} W.C.G. Ho, C.O. Heinke, \emph{A Neutron Star with a Carbon Atmosphere in the Cassiopeia A Supernova Remnant}, \emph{Nature} {\bf 462} (2009) 71.

\bibitem{H10} Heinke C.O., Ho W.C.G., \emph{Direct Observation of the Cooling of the Cassiopeia A Neutron Star}, \emph{Astrophys. J} {\bf 719} (2010) L167.

\bibitem{St} P.S. Shternin, D.G. Yakovlev, C.O. Heinke, W.C.G. Ho, D.J. Patnaude, \emph{Cooling neutron star in the Cassiopeia A supernova remnant: evidence for superfluidity in the core}, 
\emph{Mon. Not. R. Astron. Soc.} {\bf 412} (2011) L108.

\bibitem{P} D. Page, M. Prakash, J.M. Lattimer and A.W. Steiner,
\emph{Rapid cooling of the neutron star in cassiopeia A triggered by neutron superfluidity in dense matter}, 
\emph{Phys.Rev.Lett.} {\bf 106} (2011) 081101.

\bibitem{FRS76} E. Flowers, M. Ruderman, P. Sutherland, \emph{Neutrino pair emission from finite-temperature neutron superfluid and the cooling of young neutron stars }, \emph{Astrophys. J.} {\bf 205} (1976) 541.

\bibitem{YKL} D.G. Yakovlev, A.D. Kaminker and K.P. Levenfish, \emph{Neutrino emission due to Cooper pairing of nucleons
 in cooling neutron stars}, \emph{Astron. Astrophys.} {\bf 343} (1999) 650.

\bibitem{LP06} L.B. Leinson and A. P\'{e}rez, \emph{Vector current conservation and neutrino emission from singlet-paired baryons in neutron stars}, \emph{Phys. Lett.} {\bf B 638} (2006) 114.

\bibitem{L10} L.B. Leinson, \emph{Neutrino emission from triplet pairing of neutrons in neutron stars}, \emph{Phys. Rev.} {\bf C 81} (2010) 025501.

\bibitem{M88} Mayle, R., et al. \emph{Constraints on Axions from SN 1987a}, \emph{Phys. Lett.} {\bf B 203} (1988) 188.

\bibitem{M89} Mayle, R., et al. \emph{Updated Constraints on Axions from SN 1987a}, \emph{Phys. Lett.} {\bf B 219} (1989) 515.

\bibitem{L01} L. B. Leinson, \emph{Collective neutrino pair emission due to Cooper pairing of protons in superconducting neutron stars}, \emph{Nucl. Phys.} {\bf A 687} (2001) 489.

\bibitem{gs80} G. Glen, P. Sutherland, \emph{On the cooling of neutron stars}, \emph{Astrophys. J.} {\bf 239} (1980) 671.

\bibitem{Y} D.G. Yakovlev, W.C.G. Ho, P.S. Shternin, C.O. Heinke and A.Y. Potekhin, \emph{Cooling rates of neutron stars and the young neutron star in the Cassiopeia A supernova remnant}, \emph{Mon. Not. R. Astron. Soc.} {\bf 411} (2011) 1977.

\bibitem{thorne77} K.S.  Thorne, \emph{The relativistic equations of stellar structure and evolution}, \emph{Astrophys. J.} {\bf 212} (1977)  825. 

\bibitem{YLS} D.G. Yakovlev, K.P. Levenfish, Yu.A. Shibanov, \emph{Cooling of neutron stars and superfluidity in their cores}, \emph{ Phys.Usp.} {\bf 42} (1999) 737.

\bibitem{GPE}  E.H. Gudmundsson, C.J. Pethick,  R.I. Epstein, \emph{ Neutron star envelopes}, \emph{Astrophys. J.} {\bf 259} (1982) L19.

\bibitem{p06} D. Page, U. Geppert, and F. Weber, \emph{The Cooling of compact stars}, \emph{Nucl. Phys. A} {bf 777} (2006) 497.

\bibitem{K85} D.B. Kaplan, \emph{Opening the Axion Window}, \emph{Nucl. Phys.} {\bf B 260} (1985) 215.

\bibitem{S85} M. Srednicki, \emph{Axion Couplings to Matter. 1. CP Conserving Parts}, \emph{Nucl. Phys.} {\bf B 260} (1985) 689.

\bibitem{G82} J. Gasser and H. Leutwyler, \emph{Quark Masses}, \emph{Phys. Rep.} {\bf 87 } (1982) 77.
\end{thebibliography}
\end{document}